\documentclass[runningheads]{llncs}

\usepackage{amssymb}
\usepackage{latexsym}
\usepackage{amsmath}
\usepackage{mathrsfs}

\usepackage[noend]{algpseudocode}
\usepackage{algorithm}

\numberwithin{definition}{section} 
\numberwithin{lemma}{section}
\numberwithin{remark}{section}
\numberwithin{theorem}{section}
\numberwithin{proposition}{section}

\newcommand{\melemk}{\textsc{Exact \(M\)-Items Knapsack}}
\newcommand{\allocNameOne}{LPV \textsc{Allocations}}
\newcommand{\allocNameTwo}{LTV \textsc{Allocations}}

\usepackage{tikz}

\begin{document}

\title{Approximate \#Knapsack Computations to Count Semi-Fair Allocations} 

\titlerunning{An FPTAS for \#Semi-Fair Allocations}

\author{Theofilos Triommatis\inst{1} \and Aris Pagourtzis\inst{2}}
\authorrunning{T. Triommatis and A. Pagourtzis}

\institute{School of Electrical Engineering, Electronics and Computer Science, University of Liverpool, Liverpool, L69-3BX, UK\\
\email{Theofilos.Triommatis@liverpool.ac.uk}\\
\and
School of Electrical and Computer Engineering, National Technical University of Athens, Polytechnioupoli, 15780 Zografou, Athens, Greece\\
\email{pagour@cs.ntua.gr}}

\maketitle


\begin{abstract}
In this paper, we study the problem of counting the number of different knapsack solutions with a prescribed cardinality. We present an FPTAS for this problem, based on dynamic programming. We also introduce two different types of semi-fair allocations of indivisible goods between two players. By semi-fair allocations, we mean allocations that ensure that at least one of the two players will be free of envy. We study the problem of counting such allocations and we provide FPTASs for both types, by employing our FPTAS for the prescribed cardinality knapsack problem.
\end{abstract}

\section{Introduction}
We define and study three counting problems. The first of them concerns knapsack solutions with a prescribed number of items allowed in the knapsack, while the other two concern two new notions of allocations of indivisible goods among two players. We show that both our allocation notions imply a \emph{semi-fairness} property, namely that at least one of the two players is envy-free.  From a computational point of view both types of allocations are shown to be easy to satisfy, however the corresponding counting problems seem to be hard. We provide fully polynomial-time approximation schemes for all three problems that we study.  Along the way we compare our new notions of allocations to the standard notion of envy-freeness (EF)~\cite{FoleyEF} 
and show that one of them is incomparable to EF, while the other one includes all EF allocations. Note that the problem of approximate counting allocations, apart from its own interest, may serve as a basis for solving problems under \emph{uncertainty}~\cite{DBLP:journals/jcss/BuhmannGMPSW18}.

In the counting version of a decision problem that asks for the existence of a solution to the given instance we are interested in counting the number of solutions to the instance. The complexity class that characterizes counting problems with polynomial-time verifiable solutions is the well-known class \#P \cite{DBLP:journals/tcs/Valiant79} and it is known that it contains some hard counting problems. It is also known that counting problems that have NP-complete existence versions are not approximable unless NP=RP~\cite{DBLP:journals/algorithmica/DyerGGJ03}. In contrast, the class \#PE~\cite{DBLP:conf/pci/KiayiasPSZ01,DBLP:conf/mfcs/PagourtzisZ06}, consisting of counting problems in \#P that have an easy existence version,  contains many approximable counting problems.
Moreover, several well-known approximable counting problems belong to a subclass of \#PE, called TotP~\cite{DBLP:conf/mfcs/PagourtzisZ06}; in~\cite{DBLP:conf/mfcs/PagourtzisZ06} it is proven that TotP is the class that contains all the functions of \#PE that are self-reducible. Such a problem is \#Knapsack, which admits an FPTAS~\cite{DBLP:journals/siamcomp/StefankovicVV12}. Here we show, among others, that our counting problems share the property of having easy existence version, thus providing the first evidence that they admit an FPTAS.  

The first problem that we study is {\#\melemk}, a problem the optimization version of which has recently been studied~\cite{fptaskitemknapsack}. We will first present an FPTAS for this problem and then use it to obtain FPTASs for the allocation counting problems that we define in this paper. This connection could be of further interest as only few variants of \textsc{Knapsack} have been associated to allocation problems; such an example is the \textsc{Non-Linear Fractional Equality Knapsack}~\cite{DBLP:journals/eswa/YazidiJH18}.

\section{The \#\melemk\ Problem}
In this section, we define \#\melemk\ and provide an FPTAS for it. Our algorithm uses dynamic programming and builds on techniques developed in~\cite{Dyer03} and \cite{DBLP:journals/siamcomp/StefankovicVV12}. 
Firstly we will define the decision version of \#\melemk \ which is very similar to the standard \textsc{Knapsack} problem with the additional restriction that a specific number of objects should be put in the knapsack. Note that we ignore objects' values as we are interested in all feasible solutions, that is, solutions in which the sum of weights does not exceed the capacity of the knapsack.

\begin{definition}[\melemk]
\label{melemK1}
\noindent
Given the weights $\{w_1,\ldots,w_n\}$ of $n$ objects, an integer $M \in \{1,\ldots,n\}$ and a capacity $C$, is there a subset $S$ of $\{1,\ldots,n\}$ such that 
\begin{equation}
    \sum_{i \in S}{w_i} \leq C \ \ and \ \ |S|=M
\end{equation}
\end{definition}

To describe the set of feasible solutions of \melemk\ in any sub-problem we examine, we will use a function $f :\{1,\ldots,n\} \times \{1,\ldots,M\} \times \mathbb{R}^+ \rightarrow{} \mathscr{P}(\mathscr{P}(S))$ with

\begin{equation}
    f(i,m,c)= \left\{ S \subseteq \{1,\ldots,i\} : \sum_{j \in S}{w_j} \leq c \  and \  |S|=m \right\}
\end{equation}

\noindent
where $\mathscr{P}(A)$ denotes the power set of $A$.

Thus, $f(i,m,c)$ is the set of feasible knapsack solutions that use only the first $i$ objects and have exactly $m$ objects in the knapsack and total weight at most $c$. Clearly, the set of solutions to the \melemk\ problem is given by $f(n,M,C)$.

Let us now define the counting version of \melemk. 

\begin{definition}[\#\melemk]
Given the weights $\{w_1,\ldots,$ $,w_n\}$  of $n$ objects, an integer $M \in \{1,\ldots,n\}$ and a capacity $C$, how many subsets $S$ of $\{1,\ldots,n\}$ are there such that 

\begin{equation}
    \sum_{i \in S}{w_i} \leq C \ \ and \ \ |S|=M
\end{equation}
\end{definition}

\begin{remark}
Note that the solution to an instance of  \#\melemk \ is the cardinality of $f(n,M,C)$, i.e.\ $|f(n,M,C)|$.
\end{remark}

\begin{remark}
\label{obvfincr}
If the values $n$ and $M$ are fixed and $c,c' \in \mathbb{R}^+$ with $c \leq c'$ then
\begin{equation*}
    |f(n,M,c)| \leq |f(n,M,c')|
\end{equation*}
\noindent
This means that $f$ is monotone w.r.t.\ the capacity. 
\end{remark}

The \#\melemk\ problem is \#P-hard, since \textsc{\#Knapsack} can be easily reduced to it. We therefore aim at approximating it. 
Following ideas of Stefankovic \textsl{et al}~\cite{DBLP:journals/siamcomp/StefankovicVV12} we will define a function $\tau$ in order to approximate the solution of \#\melemk.

\begin{definition}
\label{deftau}
We define $\tau : \{0,\ldots,M\} \times \{0,\ldots,n\} \times \mathbb{R}^+ \longrightarrow \overline{\mathbb{R}} $ with

\begin{equation}
    \tau(m,i,a) = 
    \begin{cases}
    + \infty &  \mbox{ if } a=0 \mbox{ or } m > i, \\
    \min\left\{c\in \mathbb{R} : |f(i,m,c)| \geq a
    \right\} & \mbox{ if } a \leq \binom{i}{m} \mbox{ and } m \leq i, \\
    + \infty & \mbox{ otherwise }
    \end{cases}
\end{equation}
\end{definition}

\begin{remark}
Note that we consider as a feasible solution the one that leaves the knapsack empty, hence $\tau(0,i,1)=0$.
\end{remark}

So $\tau(m,i,a)$ represents the minimum capacity such that the number of solutions of \melemk with exactly $m$ items from  $\{1,\ldots,i\}$ is at least $a$.

We also note that $a$ should be a non negative integer, more precisely $a \in \{0, 1,\ldots,2^n\}$, but instead in the above definition we let $a \in \mathbb{R}^+$. This happens because we will approximate the number of solutions of \melemk.

Notice that with the help of function $\tau$ we can redefine the solution to an instance of \#\melemk\ as follows:
\begin{equation*}
    |f(n,M,C)| = \max\left\{ a \in \{0, 1,\ldots,2^n\}: \tau(M,n,a) \leq C \right\}
\end{equation*}

\begin{remark}
\label{obvtincr}
From Remark~\ref{obvfincr} and the definition of $\tau$ it is easy to see that for fixed $0 \leq i \leq n$, $0 \leq m \leq i$ and $a \leq a'$ we have that 
\begin{equation*}
    \tau(m,i,a) \leq \tau(m,i,a')
\end{equation*}

\noindent
This means that $\tau$ is  non-decreasing w.r.t.\ $a$.
\end{remark}

\begin{lemma}
\label{lemAnadr}
For every $i\in \{1,\ldots,n\}$, $m \in \{1,\ldots,M\}$ and $a \in \mathbb{R}^+$, $\tau$ satisfies the following recursion

\begin{equation}
\label{recMlemE}
    \tau(m,i,a) = \min_{k \in [0,1]}
    {\max{\begin{cases}
    \tau(m-1,i-1,ka)+w_i \\
    \tau \left(m,i-1,(1-k)a\right)
    \end{cases}
    }}
\end{equation}
\end{lemma}

Note that in the $i$-th step of the recursion, there are $(1-k)a$ solutions that do not contain $w_i$ and $ka$ solutions that  contain it. Furthermore we can calculate the minimum in each step if we consider every
\begin{equation*}
    k = \frac{r}{a} \ ,  \mbox{ where } r \in \mathbb{Z} \mbox{ and } 0 \leq r \leq a
\end{equation*}
By Definition~\ref{deftau} the domain of $\tau$ is $Dom(\tau) = \{1,\ldots,M\} \times \{1,\ldots,n\} \times \mathbb{R}^+$. In order to compute the exact minimum in every step of the recursion we would have to check every possible value of $r$, $0 \leq r \leq a \leq \binom{i}{m}$, thus needing in the end $\mathcal{O}(2^n)$ evaluations. We can approximate the minimum efficiently by restricting $\tau$ in $\Omega$ where 

\begin{equation*}
    \Omega = \{1,\ldots,M\} \times \{1,\ldots,n\} \times 
    \left\{ 0,1,\ldots,\lceil n\log_{Q(\varepsilon)}{2} \rceil \right\} \mbox{ and } Q(\varepsilon)=1
    +\frac{\varepsilon}{n+1}
\end{equation*}

Let $s = \lceil n \log_{Q}2 \rceil$ and $T=\left.\tau \right|_\Omega$, the restriction of $\tau$ in $\Omega$. As $T$ is a restriction of $\tau$ it must  satisfy  recursion~\ref{recMlemE}, and in particular:

\begin{equation}
\label{recT}
    T(m,i,j) = \min_{k \in [0,1]}
    {\max{\begin{cases}
    T\left(m-1,i-1,\lfloor j+\log_{Q}{k} \rfloor\right)+w_i \\
    T\left(m,i-1,\lfloor j+\log_{Q}(1-k) \rfloor\right)
    \end{cases}
    }}
\end{equation}

Now with the following algorithm we can compute $T$ efficiently and as a result we get an approximation of its optimal solution. 

\begin{algorithm}[H]
\caption{Count \melemk}
\label{algMelem}
\begin{algorithmic}[1]
\Require Integers $w_1,\ldots,w_n,C,M$ and  $\varepsilon>0$
\Ensure $(1+\varepsilon) \mbox{Approximation for \#\melemk}$
 \State Set $T[0,i,1]=0$ \ for $i \geq 0$ and \  $T[0,i,0]= \infty$ for $i \geq 0$
 \State Set $T[1,i,0]=\infty$ for $i \geq 0$ and 
 $T[1,0,j]=\infty$ for $j \geq 0$
 \State Set $T[0,i,j]=\infty$ for $i,j \geq 0$ 
 \State Set $Q = 1 + \frac{\varepsilon}{n+1}$
 \For{$m$=1 to $M$}
 \For{$i$=1 to $n$}
 \For{$j$=1 to $s$}
  \If{$\left(m>i \ or \  j>\binom{i}{m}\right)$}
   \State{$T[m,i,j]=\infty$} \vfill
   \Else
   \State{$T[m,i,j] = \min_{k \in [0,1]}
    {\max{\begin{cases}
    T\left[m-1,i-1,\lfloor j+\log_{Q}{k} \rfloor\right]+w_i \\
    T\left[m,i-1,\lfloor j+\log_{Q}(1-k) \rfloor\right]
    \end{cases}
    }}$}
   \EndIf
 \EndFor
 \EndFor
 \EndFor
 \State Set $j' = \max\{j : T[M,n,j] \leq C\}$
 \State \textbf{Return:} $Z' = Q^{j'+1}$
 \end{algorithmic}
\end{algorithm}

Now we will prove that $T$ approximates $\tau$ in the following manner

\begin{lemma} 
\label{lemmaApproxMelem}
\hfill \newline
Let $i \geq 1$, $0 \leq m \leq i$. Assume that for every $j \in \{0,\ldots,s\}$, $T[m,i-1,j]$ satisfies 

\begin{equation*}
    \tau \left(m,i-1,Q^{j-i+1} \right) \leq T[m,i-1,j] \leq \tau \left(m,i-1,Q^j \right)
\end{equation*}

\noindent
Then for all $j \in \{0,\ldots,s\}$ we have that $T[m,i,j]$ computed using~\ref{recT} satisfies:

\begin{equation*}
    \tau \left(m,i,Q^{j-i} \right) \leq T[m,i,j] \leq \tau \left(m,i,Q^j \right)
\end{equation*}

\end{lemma}

Now we are ready to prove that the output $Z'$ of Algorithm~\ref{algMelem} is a $(1+\varepsilon)$ approximation of the solution of \#\melemk.

\begin{theorem}
\label{themelemk}
 Let $Z$ be the solution of \#\melemk \ problem on an instance with $n$ items. Then for every $\varepsilon \in (0,1)$, Algorithm~\ref{algMelem} outputs $Z'$ such that
\begin{align*}
    (1-\varepsilon)Z \leq Z' \leq (1+\varepsilon)Z,
    \mbox{ \ and the algorithm runs in time \ }
    \mathcal{O} \left(\frac{n^4}{\varepsilon} \log{\frac{n}{\varepsilon}} \right)
\end{align*}

\end{theorem}

\begin{proof}
By Lemma~\ref{lemmaApproxMelem} we have for $j' = \max\{j : T[M,n,j] \leq C\}$ that the approximation $Z'$ does not underestimates $Z$ because
\begin{equation*}
    C \leq T[M,n,j'+1] \leq \tau \left( M,n,Q^{j'+1} \right)
\end{equation*}
Moreover we have at least $Q^{\left(j'-n\right)}$ solutions of \melemk\  because
\begin{equation*}
    \tau \left(M,n,Q^{\left(j'-n\right)} \right) \leq T[M,n,j'] \leq C
\end{equation*}

\begin{equation*}
    \Longrightarrow \ 
    \frac{Z'}{Z} \leq \frac{Q^{j'+1}}{Q^{j'-n}}=Q^{n+1}= \left(1+\frac{\varepsilon}{n+1} \right)^{n+1}
    \leq e^\varepsilon
\end{equation*}

This proves that the output of the algorithm satisfies the statement of the theorem.
All that is left to prove is the running time. 

The algorithm fills up a $(n \times m \times s)$ matrix with $m = \mathcal{O}(n)$. Also we have discussed above that in order to compute the minimum in recursion~(\ref{recT}) we must search all the values of a finite and discrete set $S$. More particular for every $j \in \{0,1,\ldots,s\}$, we have that $S=S_1 \cup S_2$ where $S_1 = \{Q^{-j},\ldots,Q^0\}$ and $S_2 =\{1-Q^0,\ldots,1-Q^{-j}\}$. So it will take time $\mathcal{O}(s)$ to calculate the $T[m,i,j]$ cell of the matrix. Therefore it will take time $\mathcal{O}(n m s^2)$ to fill up the matrix.

Moreover we have that $s= \lceil n \log_{Q}2 \rceil = \mathcal{O}\left(\frac{n^2}{\varepsilon}\right)$. So if the algorithm searches all the values of S in each step in order to compute the minimum of  recursion~(\ref{recT}) it will take time $\mathcal{O}\left(\frac{n^6}{\varepsilon^2}\right)$.

But from Remark~\ref{obvtincr}, we know that $\tau$ is increasing, so as $k \in [0,1]$ increases,\hfill \\
$T \left[m-1,i-1, \lfloor j+\log_{Q}{k} \rfloor \right] +w_i$  increases and $T \left[m,i-1, \lfloor j+\log_{Q}{(1-k)} \rfloor \right]$ decreases.

Now the minimum of the maximum, will be achieved for $k\in [0,1]$ with the following property: Either $k \in \{0,1\}$ or for every $k'<k$ we have
\begin{equation*}
    T \left[ m,i-1, \lfloor j+\log_{Q}{(1-k')} \rfloor \right]
    <
    T \left[m-1,i-1, \lfloor j+\log_{Q}{k'} \rfloor \right] +w_i
\end{equation*}
and for every $k'> k$
\begin{equation*}
    T \left[ m,i-1, \lfloor j+\log_{Q}{(1-k')} \rfloor \right]
    \geq
    T \left[m-1,i-1, \lfloor j+\log_{Q}{k'} \rfloor \right] +w_i
\end{equation*}
Unfortunately we can't have $S$ sorted, but we can compute $S_1$ and $S_2$ in such a way that their elements will already be in order. If we apply binary search to $S_1$ then we can find $k_1 \in [0,1]$ that satisfies the property to be the minimum of the maximum of $T$. Accordingly by applying binary search to $S_2$ we will find $k_2 \in [0,1]$ that satisfies the above property. So with this technique it takes time $\mathcal{O}(\log{s})$ to compute $T[m,i,j]$ and finally the running time of the algorithm is $\mathcal{O}\left(n  m  s  \log{s}\right)=\mathcal{O}\left( \frac{n^4}{\varepsilon} \log{\frac{n}{\varepsilon}}\right)$, concluding the proof.
\qed
\end{proof}

\section{Allocations where players value their bundle more than others do}
\label{secAlocs}
In this section we will define the problem of allocating $n$ goods between two players $A$ and $B$ in such a way that each player values its bundle more than the other player. We will assume that the $i$-th goods has value $a_i$ for $A$ and $b_i$ for $B$. More formally we have

\begin{definition}[Larger-than-swap-Player-Valuation (LPV) allocation]
\label{defAlocProb}
Given two sets $ A = \{a_i\in \mathbb{Z}^+\ : 1 \leq i \leq n \} $ and $B=\{b_i \in \mathbb{Z}^+ : 1\leq i \leq n\}$, where $n \in \mathbb{N}$, the  goal is to find a partition of   $S=\{1,\ldots,n\}$ into two sets $S_A$ and $S_B$, such that
\begin{equation}
\label{eq1DefAlocProb}
\sum_{S_A}a_i \geq \sum_{S_A}b_i  \ and \  \sum_{S_B}b_i \geq \sum_{S_B}a_i
\end{equation}
\end{definition}

In other words, the LPV allocation is a pair of bundles $(S_A,S_B)$ such that bundle $S_A$ is more valuable to $A$ than to $B$ and bundle $S_B$ is more valuable to $B$ than to $A$. 

Usually in this type of problems we are interested in fair solutions, but the interesting part is that there are many definitions of fairness. The most common notion of a fair solution is that the players should be envy free as was introduced in \cite{FoleyEF} and as the name suggests the goal is, $A$ not to envy the bundle of $B$ and vice versa. 

\begin{definition}[Envy-Free (EF) allocation]
An allocation $(S_A,S_B)$ of $n$ goods among two players $A$ and $B$, where the $i$-th good has value $a_i$ for $A$ and $b_i$ for $B$,  is called \emph{Envy Free} if
\begin{equation}
    \sum_{S_A}a_i \geq \sum_{S_B}a_i  \ and \  \sum_{S_B}b_i \geq \sum_{S_A}b_i
\end{equation}
\end{definition}

\begin{definition}[semi-Envy-Free (sEF) allocation]
For an allocation $(S_A,S_B)$ of $n$ goods among two players $A$ and $B$ wi will say that \textbf{$A$ doesn't envy $B$ or $A$ is free of envy}, and we will denote it with sEF($A$), if
\begin{equation}
    \sum_{S_A}a_i \geq \sum_{S_B}a_i
\end{equation}
\end{definition}

\begin{lemma}
In an LPV allocation at least one of the two players is free of envy. 
\end{lemma}

\begin{proof}
There are two possible cases either
\begin{equation*}
    \sum_{S_A}{b_i} \geq \sum_{S_B}{b_i} \ \  or \ \  \sum_{S_B}{b_i} \geq \sum_{S_A}{b_i}
\end{equation*}
If $\sum_{S_B}{b_i} \geq \sum_{S_A}{b_i}$ then, by definition, $B$ is free of envy.
If $\sum_{S_A}{b_i} \geq \sum_{S_B}{b_i}$ then considering the property of LPV allocation 
(\ref{eq1DefAlocProb}) we have 
\begin{equation*}
    \sum_{S_A}{a_i} \geq \sum_{S_A}{b_i} \geq \sum_{S_B}{b_i} \geq \sum_{S_B}{a_i} \Longrightarrow \ \sum_{S_A}{a_i} \geq \sum_{S_B}{a_i}
\end{equation*}
hence player $A$ is free of envy.
\qed
\end{proof}

\begin{remark}
It is easy to find an LPV allocation. We can look at all values $a_i \  and\  b_i$, if $a_i \geq b_i$ then $i \in S_A$ else $i \in S_B$. Note that there is only one LPV allocation if $a_i > b_i$  for every $1 \leq i \leq n $, namely $S_A=S, S_B=\emptyset$ (and similarly if $b_i > a_i$  for every $1 \leq i \leq n$). This means that the problem of counting LPV allocations belongs to \#PE as mentioned earlier (cf.~\cite{DBLP:conf/pci/KiayiasPSZ01,DBLP:conf/mfcs/PagourtzisZ06}.
\end{remark}

\begin{definition}[\#LPV Allocations Problem]
\label{defSAlocProb}
Given two sets $ A = \{a_i\in \mathbb{Z}^+\ : 1 \leq i \leq n \} $ and $B=\{b_i \in \mathbb{Z}^+ : 1\leq i \leq n\}$, where $n \in \mathbb{N}$, find the number of partitions of   $S=\{1,\ldots,n\}$ into two sets $S_A$ and $S_B$, such that
\begin{equation}
\label{eq1DefSAlocProb}
\sum_{S_A}a_i \geq \sum_{S_A}b_i  \ and \  \sum_{S_B}b_i \geq \sum_{S_B}a_i
\end{equation}
\end{definition}

We will now give a reduction of \#LPV \textsc{Allocations} problem to \#\melemk. This will lead to an FPTAS for the former.

\begin{lemma}
\label{lemmaAnagogiAloc}
The solution $Y$ of \#LPV \textsc{Allocations} on input $A=\{a_1,\ldots,a_n\}$ and $B=\{b_1,\ldots,b_n\}$, coincides with the sum of solutions $Z_m$ of \#\melemk \ on input $w_1,\ldots,w_n$, where $w_i=a_i-b_i+b$, and $b=\max_{1\leq i \leq n}{b_i}$, capacity $C=mb$ and exactly $m$ items in the knapsack, for $m \in \{1,\ldots,n-1\}$ (assuming w.l.o.g.\ $\sum_{i=1}^{n}{a_i} \geq \sum_{i=1}^{n}{b_i}$):
\begin{equation*}
    Y = \sum_{m=1}^{n-1}{Z_m}
\end{equation*}
    
\end{lemma}

\noindent
So an FPTAS algorithm for \#\allocNameOne\ is the following:

\begin{algorithm}[H]
\caption{Count \allocNameOne}
\label{algAloc}
\begin{algorithmic}[1]
\Require Integers $a_1,\ldots,a_n,b_1,\ldots,b_n$ and  $\varepsilon>0$
\Ensure $(1+\varepsilon) \mbox{Approximation for \#\allocNameOne}$
 \State $S_a = a_1+\cdots+a_n$
 \State $S_b = b_1+\cdots+b_n$
 \State{$Y$=0}
 \If{$S_a \geq S_b$}
    \State{$b=\max(b_1,...,b_n)$}
    \For{i=1 to n}
        \State{$w_i=a_i-b_i+b$}
    \EndFor    
    \For{$m=1$ to $n-1$}
        \State{$Y$=$Y$+Count \melemk $(w_1,...,w_n,mb,m,\varepsilon)$}
    \EndFor
 \Else
    \State{$b=\max(a_1,...,a_n)$}
    \For{$i$=1 to $n$}
        \State{$w_i=b_i-a_i+b$}
    \EndFor    
    \For{$m=1$ to $n-1$}
        \State{$Y$=$Y$+Count \melemk $(w_1,...,w_n,mb,m,\varepsilon)$}
    \EndFor
 \EndIf
 \State \textbf{Return:} $Y$
 \end{algorithmic}
\end{algorithm}

\begin{theorem}
\label{themAllocProb} 
Let Y be the solution of \#\allocNameOne \ problem. Then for every $\varepsilon \in (0,1)$, Algorithm~\ref{algAloc} outputs $Y'$ such that
\begin{align*}
    (1-\varepsilon)Y \leq Y' \leq (1+\varepsilon)Y,
    \mbox{ \ and the algorithm runs in time \ }
    \mathcal{O} \left(\frac{n^5}{\varepsilon} \log{\frac{n}{\varepsilon}} \right)
\end{align*}

\end{theorem}

\begin{proof}
Let $Z'_m$ be the output of Algorithm~\ref{algMelem} for the \#\melemk\ problem with $m$ items in the knapsack, capacity $C(m)$ (depends on $m$) and weights $w_1,w_2,\ldots,w_n$ and $Z_m$ be its exact solution. From Lemma~\ref{lemmaAnagogiAloc} we have that Algorithm~\ref{algAloc} outputs
\begin{equation*}
    Y' =  \sum_{m=1}^{n-1}{Z'_m}
    \leq 
    \sum_{m=1}^{n-1}{(1+\varepsilon)Z_m}
    =
    (1+\varepsilon) \sum_{m=1}^{n-1}{Z_m}
    =
    (1+\varepsilon) Y
\end{equation*}
Accordingly
\begin{equation*}
    Y' =  \sum_{m=1}^{n-1}{Z'_m}
    \geq 
    \sum_{m=1}^{n-1}{(1-\varepsilon)Z_m}
    =
    (1-\varepsilon) \sum_{m=1}^{n-1}{Z_m}
    =
    (1-\varepsilon) Y
\end{equation*}

As far as running time is concerned, Algorithm~\ref{algAloc} consists of simple operations that take time $\mathcal{O}(n)$ and then executes $(n-1)$ times the algorithm for \#\melemk. Since Algorithm~\ref{algMelem} runs in time $\mathcal{O}\left(\frac{n^4}{\varepsilon} \log{\frac{n}{\varepsilon}} \right)$, we obtain the claimed running time. 
\qed
\end{proof}

\section{\allocNameTwo}
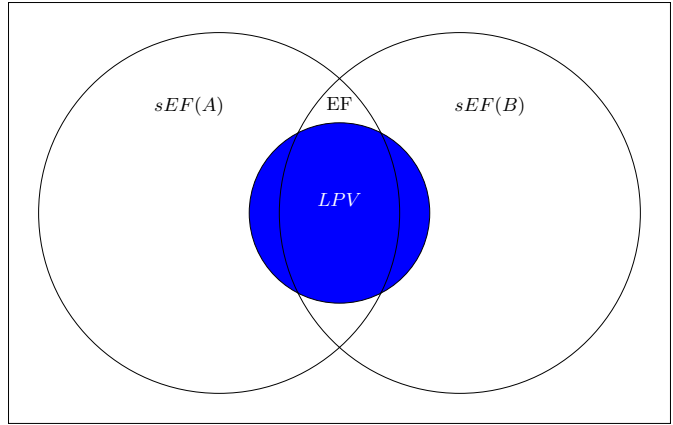
\begin{figure}
    \centering
    \scalebox{.8}{
\begin{tikzpicture}
[fill=gray]
\scope 
\fill[blue] (0.5,0) circle (1.5);
\endscope
\draw (-1.5,0) circle (3) (-2,1.5)  node [text=black,above] {$sEF(A)$}
      (2.5,0) circle (3) (3,1.5)  node [text=black,above] {$sEF(B)$}
      (0.5,0) circle (1.5) (0.5,0)  node [text=white,above] {$LPV$}
      (-5,-3.5) rectangle (6,3.5) node [text=black,above] {};
\draw  ++(20:5ex) (0.5,1.6) node[text=black, above] {EF};
\end{tikzpicture}
}
\caption{Relation of LPV to (semi-)envy-free allocations. $EF$ denotes the set of Envy-Free allocations, sEF(A) denotes the set of semi-Envy-Free allocations of $A$ (accordingly for $B$) and $LPV$ is the set of LPV allocations.}
\label{fig:LPV}
\end{figure}

In the previous section we have studied the \allocNameOne\  and we proved that in a solution of the \allocNameOne, at least one of the two players will be free of envy. However, the converse is not true, as there exist instances such as $A=\{8,4,6,5\}$ and $B=\{5,8,7,7\}$: if $A$ picks items 1 and 2 and $B$ picks items 3 and 4, i.e.\ $S_A=\{1,2\}$ and $S_B=\{3,4\}$, it is easy to confirm that the couple $(S_A, S_B)$ is an envy free solution but it doesn't satisfy Definition~\ref{defAlocProb}. The Venn diagram in Figure~\ref{fig:LPV} (appendix) visualizes the situation.

Observe that by counting LPV allocations we may miss several EF or sEF allocations. In order to capture more EF and sEF allocations, we will now define a second notion of allocations.

\begin{definition}[Larger-than-swap-Total-Valuation (LTV) allocations]
\label{defSFProb}
Given two sets $ A = \{a_i\in \mathbb{Z}^+\ : 1 \leq i \leq n \} $ and $B=\{b_i \in \mathbb{Z}^+ : 1\leq i \leq n\}$ where $n \in \mathbb{N}$, the goal is to find a partition  of $S=\{1,\ldots,n\}$ into two sets $S_A$ and $S_B$, such that $S_A\cap S_B=\emptyset$ and $S_A\cup S_B=S$ with the following property 
\begin{equation}
\label{eq1DefSFProb}
\sum_{S_A}{(a_i-b_i)} \geq \sum_{S_B}{(a_i-b_i)} 
\end{equation}
\end{definition}

\begin{proposition}
\label{PropLTVsuper}
LTV allocations contain all EF and all LPV allocations.
\end{proposition}

We can now update the Venn diagram of the allocations to include the LTV allocations, giving  a much clearer view of the inclusion relation between the allocations.

\begin{figure}
    \centering

\scalebox{.8}{
\begin{tikzpicture}[fill=gray]
\scope 
\fill[green] (0.5,0) circle (2.2);
\fill[blue] (0.5,0) circle (1.5);
\endscope
\draw (-1.5,0) circle (3) (-2,1.5)  node [text=black,above] {$sEF(A)$}
      (2.5,0) circle (3) (3,1.5)  node [text=black,above] {$sEF(B)$}
      (0.5,0) circle (1.5) (0.5,0)  node [text=white,above] {$LPV$}
      (-5,-3.5) rectangle (6,3.5) node [text=black,above] {}
      (0.5,0) circle (2.2) (0,-2) node [text=black, above] {$LTV$}
      ;
\draw  ++(20:5ex) (0.5,1.6) node[text=black, above] {EF};
\end{tikzpicture}
}
\caption{Relations between LTV, LPV and (semi) Envy Free Allocations. Notation is as in Figure~\ref{fig:LPV}, and $LTV$ denotes the set of LTV allocations.}
\label{fig:LTV}
\end{figure}
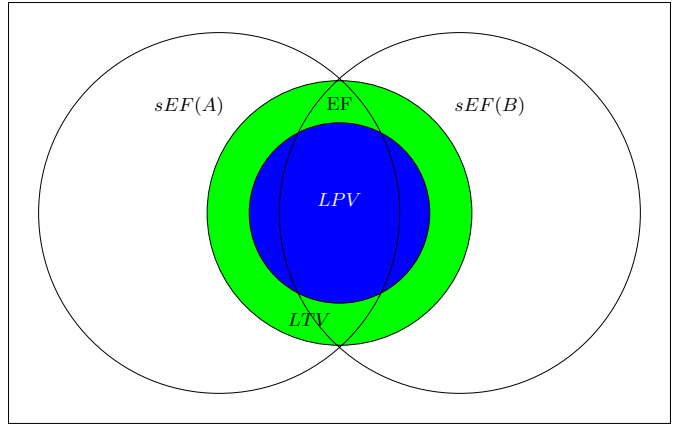

\begin{remark}
\label{rem:alloctwoeasy}
The \allocNameTwo\ problem also has some easy-to-find solutions, e.g.\ the solution that assigns to $A$ all objects that $A$ values more than $B$ and vice versa. \end{remark}

We will now define the corresponding counting problem and study its complexity.

\begin{definition}[\#\allocNameTwo]
\label{defSSFProb}
Given two sets $ A = \{a_i\in \mathbb{Z}^+\ : 1 \leq i \leq n \} $ and $B=\{b_i \in \mathbb{Z}^+ : 1\leq i \leq n\}$, where $n \in \mathbb{N}$, the goal is to find how many partitions of  $S=\{1,\ldots,n\}$ into two sets $(S_A, S_B)$ are there, such that $S_A\cap S_B=\emptyset$ and $S_A\cup S_B=S$, satisfying the following property 
\begin{equation}
\label{eqDefSSFProb}
\sum_{S_A}{(a_i-b_i)} \geq \sum_{S_B}{(a_i-b_i)}
\end{equation}
\end{definition}

By Remark~\ref{rem:alloctwoeasy} it is not unlikely that \#\allocNameTwo\ be approximable, as it belongs to \#PE (cf.~\cite{DBLP:conf/pci/KiayiasPSZ01,DBLP:conf/mfcs/PagourtzisZ06}). 
Indeed, using similar arguments to those in the proof of Lemma~\ref{lemmaAnagogiAloc} we can prove that the following algorithm is an FPTAS algorithm for \#\allocNameTwo.

\begin{algorithm}[H]
\caption{Count \allocNameTwo}
\label{algSFAloc}
\begin{algorithmic}[1]
\Require Integers $a_1,\ldots,a_n,b_1,\ldots,b_n$ and  $\varepsilon>0$
\Ensure $(1+\varepsilon) \mbox{Approximation for \#\allocNameTwo}$
 \For{i=1 to n}
    \State{$d_i=a_i-b_i$}
 \EndFor
 \State{$Sum = (d_1+\ldots+d_n)/2$}
 \State{Set $b=1-\min(d_1,...,d_n)$}
 \For{i=1 to n}
    \State{Set $w_i=a_i-b_i+b$}
 \EndFor    
 \For{$m=1$ to $n-1$}
    \State{$Y = Y + \textrm{Count} \melemk (w_1,...,w_n,Sum+mb,m,\varepsilon)$}
 \EndFor
 \State \textbf{return} $Y$
 \end{algorithmic}
\end{algorithm}

\begin{lemma}
\label{lemmaAnagogiLTV}
The solution $Y$ of \#\allocNameTwo \ on input $A=\{a_1,\ldots,a_n\}$ and $B=\{b_1,\ldots,b_n\}$, coincides with the sum of solutions $Z_m$ of \#\melemk \ on input $w_1,\ldots,w_n$, capacity $\left((\sum_{i=1}^{n}{w_i})/2 + m\beta\right)$, $\beta =\min\left\{a_i-b_i : 1\leq i \leq n \right\}$ and $w_i = a_i-b_i-\beta+1$, and exactly $m$ items in the knapsack, for $m \in \{1,\ldots,n-1\}$, that is, 
\begin{equation}
\label{eq1LemmaLTV}
    Y = \sum_{m=1}^{n-1}{Z_m}
\end{equation}
\end{lemma}

\begin{theorem}
\label{themLTV}
Let $Y$ be the exact solution of \#\allocNameTwo \ problem on some input. Then for every $\varepsilon \in (0,1)$, Algorithm~\ref{algSFAloc} outputs $Y'$ such that
\begin{align*}
    (1-\varepsilon)Y \leq Y' \leq (1+\varepsilon)Y,
    \mbox{ \ and runs in time \ }
    \mathcal{O} \left(\frac{n^5}{\varepsilon} \log{\frac{n}{\varepsilon}} \right)
\end{align*}

\end{theorem}
The proof is similar to  the proof of Theorem~\ref{themAllocProb} and is omitted. 



\section{Discussion}
We presented an FPTAS for the problem of counting feasible knapsack solutions with a specific (given) number of items; to the best of our knowledge no FTPAS has been proposed for this problem so far, despite its evident importance. We built on Dyer's dynamic programming algorithm~\cite{Dyer03}. An interesting future work would be to improve the complexity of the FPTAS by exploring dimension reduction techniques (see e.g.~\cite{MP18}).

We also defined two new notions of allocations of indivisible goods and provided FPTASs for the  counting problems associated with them by employing the above mentioned FPTAS. We leave as an open question whether our results can be extended to more than two players.

Different notions of fair allocation are examined in various papers (see, e.g.,  \cite{DBLP:conf/ijcai/AmanatidisBM18} and references therein); it would be interesting to compare these notions to our notions of LPV and LTV allocations. Moreover, we would like to see which of our techniques might be applicable to counting versions of other fair allocation problems. 

Finally, we would like to settle the complexity of counting LPV and LTV allocations either by proving \#P-hardness (as we believe is the case) or by providing polynomial-time algorithms.

\bibliographystyle{abbrv}
\bibliography{bibliography}

\begin{thebibliography}{10}

\bibitem{DBLP:conf/ijcai/AmanatidisBM18}
G.~Amanatidis, G.~Birmpas, and V.~Markakis.
\newblock Comparing approximate relaxations of envy-freeness.
\newblock In J.~Lang, editor, {\em Proceedings of the Twenty-Seventh
  International Joint Conference on Artificial Intelligence, {IJCAI} 2018, July
  13-19, 2018, Stockholm, Sweden.}, pages 42--48. ijcai.org, 2018.

\bibitem{DBLP:journals/jcss/BuhmannGMPSW18}
J.~M. Buhmann, A.~Gronskiy, M.~Mihal{\'{a}}k, T.~Pr{\"{o}}ger, R.~Sr{\'{a}}mek,
  and P.~Widmayer.
\newblock Robust optimization in the presence of uncertainty: {A} generic
  approach.
\newblock {\em J. Comput. Syst. Sci.}, 94:135--166, 2018.

\bibitem{Dyer03}
M.~E. Dyer.
\newblock Approximate counting by dynamic programming.
\newblock In L.~L. Larmore and M.~X. Goemans, editors, {\em Proceedings of the
  35th Annual {ACM} Symposium on Theory of Computing, June 9-11, 2003, San
  Diego, CA, {USA}}, pages 693--699. {ACM}, 2003.

\bibitem{DBLP:journals/algorithmica/DyerGGJ03}
M.~E. Dyer, L.~A. Goldberg, C.~S. Greenhill, and M.~Jerrum.
\newblock The relative complexity of approximate counting problems.
\newblock {\em Algorithmica}, 38(3):471--500, 2004.

\bibitem{FoleyEF}
D.~K. Foley.
\newblock Resource allocation and the public sector.
\newblock {\em Yale Econ Essays}, 7:45--98, 1967.

\bibitem{DBLP:conf/pci/KiayiasPSZ01}
A.~Kiayias, A.~Pagourtzis, K.~Sharma, and S.~Zachos.
\newblock Acceptor-definable counting classes.
\newblock In Y.~Manolopoulos, S.~Evripidou, and A.~C. Kakas, editors, {\em
  Advances in Informatics, 8th Panhellenic Conference on Informatics, {PCI}
  2001. Nicosia, Cyprus, November 8-10, 2001, Revised Selected Papers}, volume
  2563 of {\em Lecture Notes in Computer Science}, pages 453--463. Springer,
  2001.

\bibitem{fptaskitemknapsack}
W.~Li, J.~Lee, and N.~B. Shroff.
\newblock A faster {FPTAS} for knapsack problem with cardinality constraint.
\newblock {\em CoRR}, abs/1902.00919, 2019.

\bibitem{MP18}
N.~Melissinos and A.~Pagourtzis.
\newblock A faster {FPTAS} for the subset-sums ratio problem.
\newblock In L.~Wang and D.~Zhu, editors, {\em Computing and Combinatorics -
  24th International Conference, {COCOON} 2018, Qing Dao, China, July 2-4,
  2018, Proceedings}, volume 10976 of {\em Lecture Notes in Computer Science},
  pages 602--614. Springer, 2018.

\bibitem{DBLP:conf/mfcs/PagourtzisZ06}
A.~Pagourtzis and S.~Zachos.
\newblock The complexity of counting functions with easy decision version.
\newblock In R.~Kralovic and P.~Urzyczyn, editors, {\em Mathematical
  Foundations of Computer Science 2006, 31st International Symposium, {MFCS}
  2006, Star{\'{a}} Lesn{\'{a}}, Slovakia, August 28-September 1, 2006,
  Proceedings}, volume 4162 of {\em Lecture Notes in Computer Science}, pages
  741--752. Springer, 2006.

\bibitem{DBLP:journals/siamcomp/StefankovicVV12}
D.~Stefankovic, S.~Vempala, and E.~Vigoda.
\newblock A deterministic polynomial-time approximation scheme for counting
  knapsack solutions.
\newblock {\em {SIAM} J. Comput.}, 41(2):356--366, 2012.

\bibitem{DBLP:journals/tcs/Valiant79}
L.~G. Valiant.
\newblock The complexity of computing the permanent.
\newblock {\em Theor. Comput. Sci.}, 8:189--201, 1979.

\bibitem{DBLP:journals/eswa/YazidiJH18}
A.~Yazidi, T.~M. Jonassen, and E.~Herrera{-}Viedma.
\newblock An aggregation approach for solving the non-linear fractional
  equality knapsack problem.
\newblock {\em Expert Syst. Appl.}, 110:323--334, 2018.

\end{thebibliography}

\newpage

\section{Appendix}

\textbf{Proof of Lemma \ref{lemAnadr}}
\begin{proof}
Let $i \in \{1,\ldots,n\},\  m \in \{1,\ldots,M\}$ and $a \in \mathbb{R}^+$. To prove that recursion~[\ref{recMlemE}] holds we will consider that we have computed every
\begin{equation*}
    \tau(m',i',a),  \mbox{ for } m'\in \{0,\ldots,m\}, \  i' \in \{0,\ldots,i-1\} \mbox{ and } a \in [0,2^n)
\end{equation*}

\noindent
We will now prove that for some $k \in [0,1]$
\begin{equation*}
    \tau(m,i,a) \leq \min_{k \in [0,1]}
    {\max{\begin{cases}
    \tau(m-1,i-1,ka)+w_i \\
    \tau \left(m,i-1,(1-k)a\right)
    \end{cases}
    }}
\end{equation*}

Let $B(k) = \max\left\{\tau(m-1,i-1,ka)+w_i,\  \tau(m,i-1,(1-k)a)\right\}$.\smallskip

We observe that there exist at least $ka$ solutions with weights $w_1,\ldots,w_i$ and capacity no greater than $B(k)$. Accordingly there exist at least $(1-k)a$ solutions with weights $w_1,\ldots,w_{i-1}$ and capacity no greater than $B(k)$. Therefore there exist at least $a$ solutions with capacity no greater than $B(k)$. 
So we have that

\begin{equation*}
    \tau(m,i,a) \leq \min_{k \in [0,1]}{B(k)} \ \Longrightarrow
\end{equation*}

\begin{equation}
\label{eq1pRec}
    \tau(m,i,a) \leq \min_{k \in [0,1]}
    {\max{\begin{cases}
    \tau(m-1,i-1,ka)+w_i \\
    \tau \left(m,i-1,(1-k)a\right)
    \end{cases}
    }}
\end{equation}

\noindent
All that is left to prove is that

\begin{equation*}
    \tau(m,i,a) \geq \min_{k \in [0,1]}
    {\max{\begin{cases}
    \tau(m-1,i-1,ka)+w_i \\
    \tau \left(m,i-1,(1-k)a\right)
    \end{cases}
    }}
\end{equation*}

Let us consider the number of solutions to \melemk \  with weights $w_1,\ldots,w_i$, $m$ items in the knapsack and capacity $C=\tau(m,i,a)$.
For some $\beta \in [0,1]$ there are at least $\beta a$ solutions that contain the $i$-th item so

\begin{equation*}
    C=\tau(m,i,a) \geq \tau(m,i,\beta a)= \tau(m-1,i-1,\beta a)+w_i
\end{equation*}

Moreover there are also $(1-\beta)$ solutions that do not contain the $i$-th item so $\tau(m,i,(1-\beta) a) \equiv \tau\left(m,i-1,(1-\beta)a\right)$. For these solutions we have that 
\begin{equation*}
    C=\tau(m,i,a) \geq \tau(m,i,(1-\beta) a) = \tau\left(m,i-1,(1-\beta)a\right)
\end{equation*}

It is easy to check that the above inequalities are true if we bear in mind Remark~\ref{obvtincr}. Now we have that for some $\beta \in [0,1]$

\begin{equation*}
    C = \tau(m,i,a) \geq \max\{\tau(m-1,i-1,\beta a)+w_i, \ \tau(m,i-1,(1-\beta) a) \}
\end{equation*}

Hence

\begin{equation}
\label{eq2pRec}
    \tau(m,i,a) \geq \min_{k \in [0,1]}
    {\max{\begin{cases}
    \tau(m-1,i-1,ka)+w_i \\
    \tau \left(m,i-1,(1-k)a\right)
    \end{cases}
    }}
\end{equation}

From equations (\ref{eq1pRec}) and (\ref{eq2pRec}) we obtain (\ref{recMlemE}). \qed
\end{proof}

\medskip
\noindent
\textbf{Proof of Lemma \ref{lemmaApproxMelem}}
\begin{proof}
Let $i \geq 0$,  $0 \leq m \leq i$. By the assumption of the lemma and remark~\ref{obvtincr} we have that 

\begin{equation}
\label{eq1Ppros}
\begin{split}
    T \left[m-1,i-1,\lfloor j+\log_{Q}{k} \rfloor \right]
    &\geq 
    \tau \left(m-1,i-1,Q^{\left(\lfloor j+\log_{Q}{k} \rfloor-(i-1) \right)} \right)
    \\ \hfill  
    &\geq
    \tau \left(m-1,i-1,kQ^{j-i} \right)
\end{split}
\end{equation}

\begin{equation}
\label{eq2Ppros}
\begin{split}
    T \left[m,i-1,\lfloor j+\log_{Q}{(1-k)} \rfloor \right]
    &\geq 
    \tau \left(m,i-1,Q^{\left(\lfloor j+\log_{Q}{(1-k)} \rfloor-(i-1) \right)} \right)
    \\ \hfill  
    &\geq
    \tau \left(m,i-1,(1-k)Q^{j-i} \right)
\end{split}
\end{equation}

From inequalities~\ref{eq1Ppros} and~\ref{eq2Ppros} we have

\begin{equation*}
\begin{split}
    T[m,i,j] &= \min_{k \in [0,1]}{
    \max{ 
    \begin{cases}
     T \left[m-1,i-1,\lfloor j+\log_{Q}{k} \rfloor \right] + w_i \\
     T \left[m,i-1,\lfloor j+\log_{Q}{(1-k)} \rfloor \right]
    \end{cases} }}
    \\ \hfill
    &\geq
    \min_{k \in [0,1]}{
    \max{ 
    \begin{cases}
     \tau \left(m-1,i-1,kQ^{j-i} \right)+w_i \\
     \tau \left(m,i-1,(1-k)Q^{j-i} \right)
    \end{cases} }}
    \\ \hfill
    &=
    \tau \left(m,i,Q^{(j-i)} \right) \ 
    \Longrightarrow \  T[m,i,j] \geq \tau \left(m,i,Q^{(j-i)} \right)
\end{split}
\end{equation*}

Using the same arguments as above, we prove the upper bound of the lemma.

\begin{equation}
\label{eq3Ppros}
\begin{split}
    T \left[m-1,i-1,\lfloor j+\log_{Q}{k} \rfloor \right]
    &\leq 
    \tau \left(m-1,i-1,Q^{\left(\lfloor j+\log_{Q}{k} \rfloor \right)} \right)
    \\ \hfill  
    &\leq
    \tau \left(m-1,i-1,kQ^{j} \right)
\end{split}
\end{equation}

\begin{equation}
\label{eq4Ppros}
\begin{split}
    T \left[m,i-1,\lfloor j+\log_{Q}{(1-k)} \rfloor \right]
    &\leq 
    \tau \left(m,i-1,Q^{\left(\lfloor j+\log_{Q}{(1-k)} \rfloor \right)} \right)
    \\ \hfill  
    &\leq
    \tau \left(m,i-1,(1-k)Q^{j} \right)
\end{split}
\end{equation}

Accordingly from inequalities~\ref{eq3Ppros} and~\ref{eq4Ppros} we have

\begin{equation*}
\begin{split}
    T[m,i,j] &= \min_{k \in [0,1]}{
    \max{ 
    \begin{cases}
     T \left[m-1,i-1,\lfloor j+\log_{Q}{k} \rfloor \right] + w_i \\
     T \left[m,i-1,\lfloor j+\log_{Q}{(1-k)} \rfloor \right]
    \end{cases} }}
    \\ \hfill
    &\leq
    \min_{k \in [0,1]}{
    \max{ 
    \begin{cases}
     \tau \left(m-1,i-1,kQ^{j} \right)+w_i \\
     \tau \left(m,i-1,(1-k)Q^{j} \right)
    \end{cases} }}
    \\ \hfill
    &=
    \tau \left(m,i,Q^{(j)} \right) \ 
    \Longrightarrow \  T[m,i,j] \leq \tau \left(m,i,Q^{(j)} \right)
\end{split}
\end{equation*} 
\end{proof}

\noindent
\textbf{Proof of Lemma \ref{lemmaAnagogiAloc}}
\begin{proof}
Without loss of generality we consider 
\begin{equation}
\label{eq1anag}
    \sum_{i=1}^{n}{a_i} \geq \sum_{i=1}^{n}{b_i}
\end{equation}

(if not, we can exchange $A$ and $B$.)

Now, if there exists a set $S_1$ such that 
\begin{equation}
\label{eq2anag}
    \sum_{S_1}{b_i} \geq \sum_{S_1}{a_i}  
    \Longleftrightarrow
    -\sum_{S_1}{a_i} \geq - \sum_{S_1}{b_i} ,
\end{equation}
by adding inequalities~[\ref{eq1anag}] and~[\ref{eq2anag}] we have that
\begin{equation*}
   \sum_{i=1}^{n}{a_i} -\sum_{S_1}{a_i} 
   \geq
   \sum_{i=1}^{n}{b_i} - \sum_{S_1}{b_i}
   \Longleftrightarrow
   \sum_{S \backslash S_1}{a_i} 
   \geq 
   \sum_{S \backslash S_1}{b_i}
\end{equation*}
This means that if a set $S_1$ satisfies inequality~[\ref{eq2anag}] then $(S\backslash S_1, S_1)$  is an LPV allocation. It is now obvious that instead of counting all pairs $(S_A,S_B)$ that are LPV allocations, we may just count all sets $S_1 \subseteq S$ for which

\begin{equation*}
    \sum_{S_1}{b_i} \geq \sum_{S_1}{a_i}  
    \Longleftrightarrow
    \sum_{S_1}{(a_i-b_i)} \leq 0
\end{equation*}

\begin{equation}
\label{eq3anag}
    \Longrightarrow
    \sum_{S_1}{(a_i-b_i)}+|S_1|b \leq |S_1|b
    \Leftrightarrow
    \sum_{S_1}{(a_i-b_i+b)} \leq |S_1|b
\end{equation}
Setting 
\begin{equation*}
b = \max_{1\leq i \leq n}{b_i}    
\end{equation*}
for every $i \in \{1,...,n\}$ and $S_1 \subseteq S$ we have that

\begin{equation}
\label{eq4anag}
    \sum_{S_1}{(a_i-b_i+b)} \geq 0
\end{equation}
From~[\ref{eq3anag}] and~[\ref{eq4anag}] we have

\begin{equation}
\label{eq5anag}
    0 \leq \sum_{S_1}{(a_i-b_i+b)}
    \leq |S_1|b
\end{equation}
If we now set $w_i := a_i-b_i+b$ and $C:=|S_1|b$ then inequality~[\ref{eq5anag}] is equivalent to 
\begin{equation*}
    0 \leq \sum_{S_1}{w_i} \leq C
\end{equation*}
In conclusion, to find the number of LPV allocations we need to find, for each $1\leq m\leq n-1$, how many sets $S_1 \subseteq S$ are there of cardinality $|S_1|=m$ satisfying inequality~[\ref{eq5anag}] for $C=mb$. 
Summing up all these numbers we will get our answer.
\qed
\end{proof}

\noindent
\textbf{Proof of Proposition \ref{PropLTVsuper}}
\begin{proof}
Let us consider $S_A$, $S_B$ such that $(S_A,S_B)$ is envy free, so
\begin{equation}
    \sum_{S_A}{a_i} \geq \sum_{S_B}{a_i}
    \mbox{ and }
    \sum_{S_B}{b_i} \geq \sum_{S_A}{b_i}
\end{equation}
By adding the two inequalities we have 
\begin{equation}
\label{eq1PropSF}
    \sum_{S_A}{a_i}+\sum_{S_B}{b_i} 
    \geq 
    \sum_{S_B}{a_i}+\sum_{S_A}{b_i}
    \Longrightarrow
    \sum_{S_A}{a_i}-\sum_{S_A}{b_i} 
    \geq 
    \sum_{S_B}{a_i}-\sum_{S_B}{b_i}
\end{equation}
Thus yielding inequality (\ref{eq1DefSFProb}). 

Now let us consider $S_A$ and $S_B$ to be an LPV allocation, so they will satisfy the following
\begin{equation}
\sum_{S_A}a_i \geq \sum_{S_A}b_i  \ and \  \sum_{S_B}b_i \geq \sum_{S_B}a_i
\end{equation}
Again if we add the two inequalities we get (\ref{eq1PropSF}) and then (\ref{eq1DefSFProb}). \qed
\end{proof}

\noindent
\textbf{Proof of Lemma \ref{lemmaAnagogiLTV}}
\begin{proof}
If there exists a set $S_1$ such that 
\begin{equation}
\label{eq1anagLTV}
    \sum_{i \in S_1}{a_i-b_i} \leq \frac{\sum_{i=1}^{n}{a_i-b_i}}{2}  
\end{equation}
Then for its complement $S\backslash S_1$ we will have that
\begin{equation}
\label{eq2anagLTV}
    \sum_{i \in S \backslash S_1} {a_i-b_i} \geq \frac{\sum_{i=1}^{n}{a_i-b_i}}{2}
\end{equation}
From inequalities (\ref{eq1anagLTV}) and (\ref{eq2anagLTV}) we have that
\begin{equation*}
   \sum_{i \in S_1}{a_i-b_i} 
   \leq
   \frac{\sum_{i=1}^{n}{a_i-b_i}}{2}
   \leq
   \sum_{i \in S \backslash S_1} {a_i-b_i}
\end{equation*}
This means that if a set $S_1$ satisfies inequality~[\ref{eq1anagLTV}] then $(S_1, S\backslash S_1)$  is a solution for the \#\allocNameTwo. It is now obvious that instead of counting all pairs $(S_1,S_2)$ which are solutions to the problem, we may count all $S_1 \subseteq S$ that satisfy (\ref{eq1anagLTV}). One problem that arises is that if we were to use Algorithm~\ref{algMelem} we would need all $w_i$ to be non negative, so we can't use  $w_i=a_i-b_i$. To avoid having $w_i \leq 0$ we do the following trick.
Let $\beta =\min\left\{a_i-b_i : 1\leq i \leq n \right\}$ and $w_i = a_i-b_i-\beta+1$ for $i \in \{1,...,n\}$. Now we can confirm that for $i \in \{1,...,n\}$
\begin{equation*}
    a_i-b_i \geq \beta
    \Longleftrightarrow
    a_i-b_i-\beta \geq 0
    \Longrightarrow
    a_i-b_i-\beta+1 > 0
\end{equation*}
So for every $i \in \{1,...n\}$, $w_i>0$. To count the solutions of \#\allocNameTwo\ we have to count all $S_1$ that satisfy (\ref{eq1anagLTV}) which is equivalent to counting all $S_1$ that satisfy
\begin{equation}
\label{eq3anagLTV}
    \sum_{i \in S_1}{(a_i-b_i-\beta+1)} 
    \leq 
    \frac{\sum_{i=1}^{n}{a_i-b_i}}{2}+|S_1|(1-\beta)  
\end{equation}
If we name $C := |S_1|(1-\beta)$ then inequality (\ref{eq3anagLTV}) is equivalent to 
\begin{equation*}
    0 \leq \sum_{S_1}{w_i} \leq C
\end{equation*}
Therefore, to find the solution of \#\allocNameTwo\ we need to find how many $S_1 \subseteq S$ are there with cardinality $|S_1|$, such that they satisfy inequality~[\ref{eq3anagLTV}].
If we search for every possible $|S_1|$, with  $1 \leq |S_1| \leq n-1$ and sum up all these solutions we will get (\ref{eq1LemmaLTV}). \qed
\end{proof}

\end{document}